\documentclass[prl,twocolumn,aps,showpacs,amsmath,amssymb]{revtex4}
\usepackage{graphicx}
\usepackage{dcolumn}
\usepackage{bm}

\begin{document}

\title{Spontaneous mass current and textures of
$p$-wave superfluids of trapped Fermionic atom gases
 at rest and under rotation}

\author{Y. Tsutsumi}
\affiliation{Department of Physics, Okayama University,
Okayama 700-8530, Japan}
\author{K. Machida}
\affiliation{Department of Physics, Okayama University,
Okayama 700-8530, Japan}
\date{\today}

\begin{abstract}
It is found theoretically based on the Ginzburg-Landau framework
that  $p$-wave superfluids  of neutral atom gases in three dimension harmonic traps
exhibit spontaneous mass current at rest, whose direction depends on trap geometry.
Under rotation various types of the order parameter textures are stabilized, 
including Mermin-Ho and Anderson-Toulouse-Chechetkin vortices.
In a cigar shape trap spontaneous current flows longitudial to the rotation axis
and thus perpendicular to the ordinary rotational current.
These features, spontaneous  mass current at rest and texture formation, can be used as diagnoses
for $p$-wave superfluidity.
\end{abstract}


\maketitle

Superfluids with multi-component order parameter (OP) form
a unique physics branch, including from
spinor BEC\cite{dan,stenger,ohmi,ho} in cold Bosonic atom gases with $^{23}$Na or $^{87}$Rb, strong interacting 
Fermionic liquid $^3$He atoms\cite{leggett,wolfle,volovik,salomaa}, some of the
heavy Fermion superconductors, such as UPt$_3$\cite{machida,sauls}.
It is further extending to color superconductivity in dense 
quark-gluon plasmas in high energy physics\cite{buballa,alford}.
This is unique because one can expect rich
topological defect structures or vortices and explore
a new phase of matter interesting from a fundamental physics view point.
Some of vortices, which are able to accommodate Majorana zero mode at 
a core may be useful in quantum computing\cite{sarma}.

Recently there have been much attention, in this respect, on 
$p$-wave resonance superfluidity made by Fermionic alkaline atom gases, 
such as $^6$Li\cite{zheng,schunck} and $^{40}$K\cite{jin1,jin2,jin3} 
both experimentally and theoretically\cite{yip,leo}.
Since experiments for achieving $p$-wave resonance superfluidity are steadily 
progressing\cite{inada}, it might be a good timing and also necessary to 
consider the generic properties of $p$-wave superfluidity both at rest and 
under rotation in order to help detecting its superfluidity
where non-trivial topological structures, or vortices are a hallmark 
to identify it. Namely, the textural structure of the three component order 
parameters spanned by three basis functions $p_x$, $p_y$ and $p_z$.
The order parameter space consisting of three components
is analogous to that of 
the superfluid $^3$He, in particular, the A phase\cite{wolfle,leggett,volovik}
where the OP is described by a tensor $A_{\mu,i}=d_{\mu}A_i$
($\mu$, $i=x,y,z$). The $d$-vector ($A_i$) describes the spin (orbital)
state of a Cooper pair. Since in the $p$-wave Feshbach resonance superfluid  
under a magnetic sweeping the spin  degrees of freedom is frozen,
only the orbital degrees of freedom is active, namely
the OP is characterized only by $A_i$.
In a sense our $p$-wave superfluid is analogous to the ``spinless'' superfluid $^3$He-A phase\cite{tsutsumi}.

A critical difference between superfluid $^3$He and $p$-wave resonance superfluid of 
atom gases lies in the boundary condition: Atom gases
are confined three-dimensionally (3D) by a harmonic 
trap, which is easily controlled, resulting in various shapes, such as
cigar or pancake shapes. As we will see soon, the trapping potential gives rise
to an important handle to control the the 3D texture structure.
Indeed, the 3D trapping structure constrains possible
textures as a whole. This feature is usually absent in superfluid $^3$He where 2D textures,
such as Mermin-Ho\cite{MH} or Anderson-Toulouse-Chechetkin\cite{AT} are discussed.
They are essentially 2D textures. Here we are interested in finding a truly 3D object. 

The dipole-dipole interaction between two alkaline atoms
acts to split the relative orbital state for two particles, depending upon 
the projections of the orbital angular momentum, $m_l=\pm 1$ and $m_l=0$.
This results in breaking the degeneracy among $A_x\pm iA_y$
and $A_z$, or $p_x\pm ip_y$ and $p_z$ are non-generate.
This splitting is estimated to be large for $^{40}$K by Cheng and Yip\cite{yip}
evidenced by the clear difference in the Fershbach resonance
magnetic fields (the splitting field=0.47$\pm$0.08G)\cite{jin2}.
For $^6$Li it could be small where there is no clear observation
of the resonance splitting at a magnetic field $H=158.5(7) $G in Univ. Tokyo experiment so far\cite{inada}.
In this paper we take its splitting as a parameter to extract
the generic features of the $p$-wave superfluids.
Our purposes are to find the possible 3D texture formed by the $\vec l$-vector defined shortly
in 3D harmonic trap and help identifying the $p$-wave superfluidity.

The superfluid condensate is described as 
$\Delta=A_xp_x+A_yp_y+A_zp_z=A_+p_++A_-p_-+A_0p_z$
with $p_{\pm}=\mp(p_x\pm ip_y)/{\sqrt 2}$, $A_{\pm}=\mp(A_x\mp iA_y)/{\sqrt 2}$,
$A_0=A_z$.  $A_{\pm}$  and $A_0$ are the three OP's for the $p$-wave superfluid.
The $\vec l$-vector is defined as $l_x({\bf r})={ \sqrt 2} Re \{ (A_+ + A_- )A_0^*\}/| \Delta |^2,
l_y({\bf r})={ \sqrt 2} Im \{ (-A_+ + A_- )A_0^*\}/| \Delta |^2, l_z({\bf r})=( |A_+|^2 - |A_-|^2)/ | \Delta |^2$.
This $\vec l$-vector now fully characterizes the $p$-wave superfluid and its spatial variation is called 
the $\vec l$-vector texture or simply texture.

Here we employ the Ginzburg-Landau (GL) framework, which is free
from the microscopic details and which is able to lead to generic topological 
structure of the problem. Namely, we start with the GL free energy functional which is expanded in terms of the OP's $A_{\pm}({\bf r})$  and $A_0({\bf r})$ up to the fourth oder,

$$f=f_{grad}+f_{bulk}+f_{harmonic}+f_c$$
$$f_{grad}=K_1(\partial_i^* A_j^* )(\partial_i A_j )+K_2(\partial_i^* A_j^* )(\partial_j A_i )+K_3(\partial_i^* A_i^* )(\partial_j A_j )$$
$$f_{bulk}=-\alpha_0 (1-t_i) A_i^* A_i + \beta_{24} A_i^* A_i A_j^* A_j + \beta_3 A_i^* A_i^* A_j A_j $$
$$f_{harmonic}={1\over 2}m\omega_{\perp}^2(\rho^2+\lambda^2z^2)|A_i|^2$$

\noindent
where $t_i=T/T_{ci}$ ($T_{ci}$ is the transition temperature for $i$-component),
$\partial_i=\nabla_i-i({\vec \Omega}\times {\vec r})_i $, and 
$\rho^2=x^2+y^2$. We consider ${\vec \Omega}\parallel z$.
The anisotropy of the harmonic trap is expressed as $\lambda\equiv \omega_{z}/\omega_{\perp}$. 
The GL parameters $\alpha_0$, $\beta_{24}=\beta_2+\beta_4$, $\beta_3$ and $K_1=K_2=K_3=K$ are taken as those 
estimated   by the weak coupling approximation, assuming the
Fermi sphere\cite{wolfle}:
$\alpha_0=N(0)/3$, $\beta_2=\beta_3=\beta_4=7\zeta(3)N(0)/120(\pi k_BT_c)^2$ and 
$K=7\zeta(3)N(0)(\hbar v_F)^2/240(\pi k_BT_c)^2$ where $N(0)$ is the density of states at the Fermi level,
and $v_F$ is the Fermi velocity.
The weak coupling approximation should be a good guide to understand the generic properties of the 
$p$-wave superfluids of atom gases because it has applied successfully even to liquid $^3$He of strong interacting
Fermions with only additional small strong corrections\cite{wolfle}.

It is interesting to notice that the centrifugal 
potential leads to non-trivial form
$f_c=-{1\over 2}m\Omega^2\rho^2(|A_{\rho}|^2+3|A_{\theta}|^2|+A_{z}|^2)$
with $A_{\rho}=A_x \cos{\theta} + A_y \sin{\theta}$, $A_{\theta}=-A_x \sin{\theta} + A_y \cos{\theta}$ because the OP label implies the orbital angular momentum, a feature absent in the spinor BEC.
The extra factor 3 in the above becomes important when evaluating the critical angular velocity $\Omega_{cr}$ above which the superfluid turns to be normal. That is, $\Omega_{cr}=\omega_{\perp}/\sqrt 3$
is greatly reduced from the usual case ($\Omega_{cr}=\omega_{\perp}$).
The trapping potential $f_{harmonic}$
acts as lowering the transition temperatures.
As mentioned, the dipole interaction causes the splitting of the transition 
temperatures into the two groups $T_{cx}=T_{cy}$ and $T_{cz}$.
We introduce $\alpha=T_{cx}/T_{cz}$ $(0\le\alpha\le1)$ which indicates the degrees of the broken symmetry of the system, that is, the three components are completely degenerate for $\alpha=1$. When $\alpha\rightarrow 0$, the one component $A_0(\neq0)$
or scaler superfluid tends to be realized.

Before going into the confined system, we first consider an infinite system.
The phase diagram shown in Fig.1 consists of the three phases:
The lower A phase is described by a chiral OP $A_0+iA_j$ ($j=x$ or $y$)
where the time reversal symmetry is broken. The high temperature B phase is 
described by $A_0$ which is the one component scaler superfluid.
In the following we examine the A phase in the confined geometries.

\begin{figure}
\includegraphics[width=6cm]{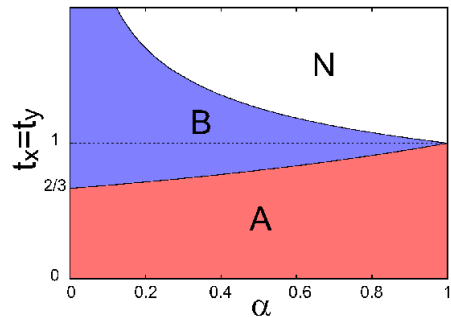}
\caption{(Color online) 
Phase diagram of the chiral $p$-wave pairing state in infinite system
in temperature ($t_x=t_y$) versus the anisotropy $\alpha\equiv T_{cx}/T_{cz}$.
N: normal state, B: single component state $A_0$ and B: the chiral state $A_0+iA_j$
($j=x$ or $y$). 
}
\end{figure}

We have found stationary solutions by numerically solving the variational
equations: $\delta F/\delta A_i({\bf r})=0$ $(i=\pm,0)$ in three dimensions.
For a cigar shape trap,  $80\times80\times 120$ meshes  are
taken with the cloud sizes $R_x=R_y=10\xi_0$ and $R_z=50\xi_0$.
The coherent length $\xi_0\equiv 0.649\hbar v_F/2\pi kT_c$. 
For a pancake shape trap,  $100\times100\times 80$ meshes  are
taken with the cloud sizes $R_x=R_y=30\xi_0$ and $R_z=10\xi_0$.
We started various initial configurations, which are uniform,
include singular or non-singular vortex, to make sure that 
the resulting texture is energy-minimum.
Throughout the paper the temperature $t_x=t_y=0.4$ and $\alpha=0.9$ are fixed
for the cigar trap ($\lambda=0.2$). For the pancake trap ($\lambda=3.0$)
$t_x=t_y=0.5$ and $\alpha=0.95$ are fixed

It is noted that in our problem the boundary condition is essential in 
determining a stable texture.
In superfluid $^3$He-A phase the $\vec l$-vector is always perpendicular to the 
hard wall so that the perpendicular particle motion is suppressed.
In other words, the point nodes situated to the
$\vec l$-vector direction touch the hard wall so as to minimize the condensation loss 
at the boundary\cite{wolfle}. Now in our harmonic trap where the condensation density 
gradually decreases towards the outer region, the $\vec l$-vector tends to 
align parallel to the circumference. This orientation is advantageous because 
the condensation energy is maximally gained by letting the point nodes move out from the system.
We can expect quite different situations for cigar and pancake shapes, 
which indeed leads to quite distinctive 3D textures.

We start out with the stable texture at rest for a cigar shape trap with $\lambda=0.2$.
At $\Omega=0$ the $\vec l$-vector is shown in Fig.2.
Figure 2(a) displays the amplitude distribution of the $\vec l$-vector.
It is seen that  the amplitude $|\vec l |$ is maximal in the central region and towards the outer regions $|\vec l | $
becomes smaller gradually.
At the top and bottom ends the polar state is realized where the $\vec l$-vector
vanishes.
The three cross sections are shown in Figs.2 (b), (c) and (d).
In Fig.2(c) which corresponds to the middle cross section the $\vec l$-vectors lie in 
the $x$-$y$ plane, showing  a stream line type pattern in which the $\vec l$-vectors
follow the circumference as if the water streams along the circular boundary.  In the outside of the condensate, 
a unseen sink and source of the $\vec l$-vector exist so that two imaginative focal points
situated outside. The left (right) one corresponds to a source (sink) where the $\vec l$-vector appears (disappears). 
This stream line like texture is contrasted with the so-called Pan-Am type texture 
in superfluid $^3$He A phase\cite{wolfle} where the $\vec l$-vectors tends to point
perpendicular to the wall due to the boundary condition.
In the upper (Fig. 2(b)) and lower (Fig. 2(d))
cross sections the stream line type textures maintain, but $l_z$ component appears
additionally.

\begin{figure}
\includegraphics[width=8cm]{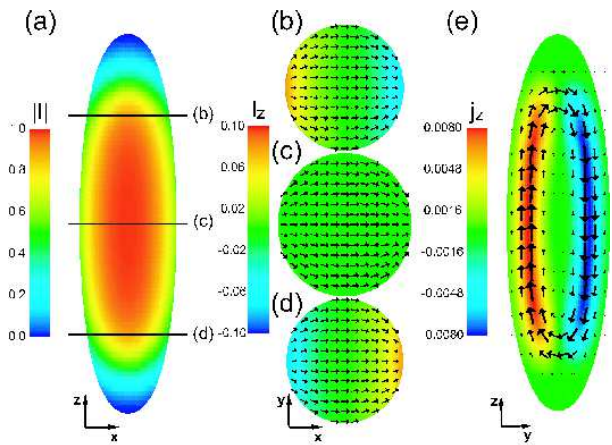}
\caption{(Color online) 
Stable texture at rest for the cigar trap $\lambda=0.2$.
(a) Distribution of $|l|$ in $z$-$x$ plane. (b)-(d) Three cross sections shown in (a)
for $l_z$  (color) and $l_x$ and $l_y$ components (arrows).
(e) Spontaneous circulating current flows in $z$-$y$ plane along the $z$ direction.
}
\end{figure}

\begin{figure}
\includegraphics[width=8cm]{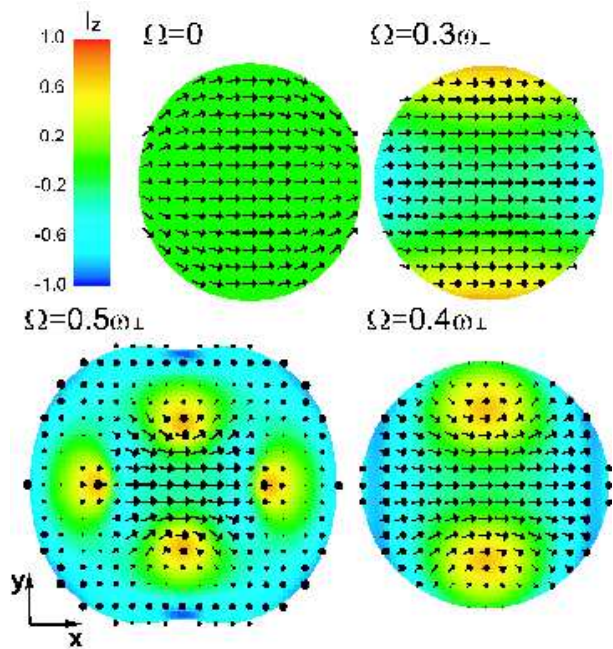}
\caption{(Color online) 
Texture change with rotation speeds $\Omega$ relative to the in-plane trap frequency $\omega_{\perp}$. 
Central cross sections are displyed. $l_z$  (color) and $l_x$ and $l_y$ components (arrows).
$\Omega=0$ corresponds to (c) in Fig.1.  As increaing $\Omega$ $\vec l$-vectors acquire the $z$ component
(color changes). At $\Omega=0.4\omega_{\perp}$ two Mermin-Ho vortices enter from $y$ direction seen as
yellow objects. At $\Omega=0.5\omega_{\perp}$ four MH's are visible and the condensate expands and distorts.
}
\end{figure}

The associated mass current structure is depicted in Fig.2(e).
The $j_z$ component shows the circulation current along the $z$ axis.
Since $\Omega=0$, this circulating mass current is spontaneously generated.
This non-trivial mass flow is explained in terms of the so-called bending current:
${\vec j}\propto {\vec \nabla}\times{\vec l}$.
This is not usual direct current due to the OP phase modulation because 
of $\Omega=0$. There is no phase twisting here.
This is due to the $\vec l$-vector bending. In the central cross section in  Fig.2(c) the 
$\vec l$-vector in-plane bending $({\vec \nabla}\times{\vec l})_z$ produces the perpendicular current $j_z$.
However, at the upper (lower) plane Fig.2(b) (Fig.2(d)) the mass current 
acquires the $j_x$ and $j_y$ components because of the non-vanishing $l_z$
component appearing there.
Therefore, the perpendicular current at the center bends around so that 
the total mass is conserved as required.
It is clear from Fig.2(e) that the mass current circulates perpetually
along the $z$ direction parallel to the long axis of the trap. This result is non-trivial and
remarkable to manifest itself the topological nature of the ${\vec l}$ texture.

Under rotation, the above 3D texture deforms continuously and smoothly so as to accommodate 
the in-plane circular direct current of the usual type.
It is seen from Fig. 3 where the central cross sections of the cigar under rotation
are displayed that as rotation increases, 
(1) the $\vec l$-vectors completely in the $x$-$y$ plane  pointing $x$ direction
acquire the $z$ component as seen by color change from green to blue. 
(2) Above a certain rotation ($\Omega=0.4\omega_{\perp}$) a pair of Mermin-Ho (MH) like vortex\cite{MH}
enters from the $y$ direction,
where at the core the  $\vec l$-vector pointing to $z$ direction flares out circularly
and at far sites it lies almost in $x$-$y$ plane. This vortex is nothing but Mermin-Ho 
structure.
(3) Upon further increasing rotation ($\Omega=0.5\omega_{\perp}$)
two pairs of the MH vortex appears.
(4) Gradually and concurrently the background $\vec l$-vectors
point to the negative $z$ direction.
(5) Moreover, the condensate profile itself deforms
and deviates  from circular. Simultaneously it spreads out due
to higher rotation, seen as extended area for $\Omega=0.5\omega_{\perp}$ case.

Figure 4 shows a different view of Fig.3 where the $z$-$x$ cross-section is displayed.
At $\Omega=0$ it is clearly seen that all $\vec l$-vectors point to the $x$ direction.
As $\Omega$ increases, the down-ward $l_z$ component appears, which is responsible
for the counter-clock-wise circular mass current.
At the top and bottom ends the polar core vortices appear as indicated by star marks.
The $\Omega=0.5\omega_{\perp}$ case shows the side view of the MH vortex:
The red lines indicated by arrows correspond to the cores of MH, which bend outward,
inducing the profile modification of the condensate near the arrows.

\begin{figure}
\includegraphics[width=8cm]{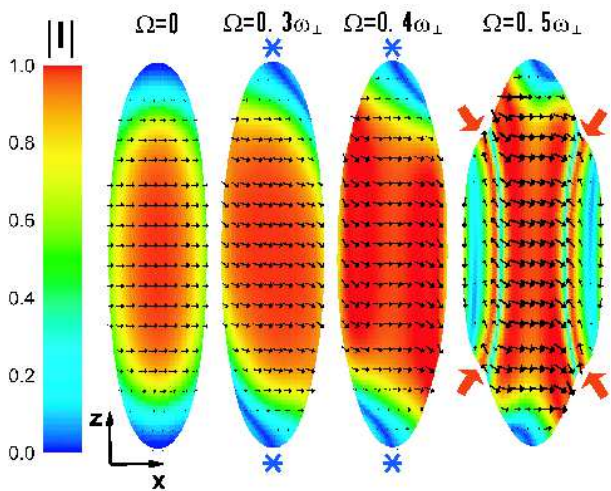}
\caption{(Color online) 
Corresponding $z$-$x$ cross sections to Fig.3.
The $\vec l$-vectors are almost in $x$-$y$ plane at $\Omega=0$.
Under rotation two polar core vortices shown in stars appear.
At $\Omega=0.5\omega_{\perp}$ two of the Mermin-Ho vortices out of four
can be seen as indicated by red arrows.
}
\end{figure}

\begin{figure}
\includegraphics[width=9cm]{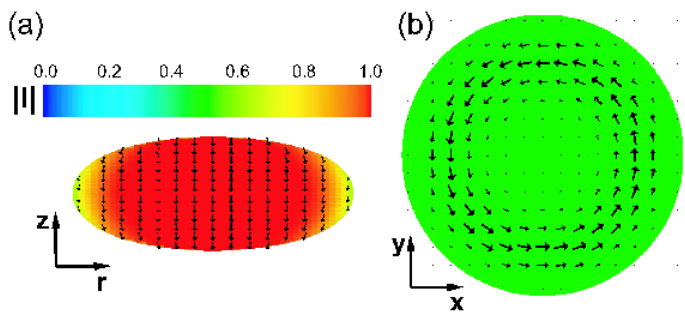}
\caption{(Color online) 
Stable texture in the pancake trap with $\lambda=3.0$ at rest.
(a) The $\vec l$-vector pattern in $z$-$r$ plane. It
is axis-symmetric around $z$.
(b) Spontaneous mass current in $x$-$y$ plane at $z=0$
}
\end{figure}

Let us now discuss the pancake shape trap.
From Fig.5 where the resulting $\vec l$-vector texture (Fig.5(a))
and associated in-plane mass current (Fig.5(b)) at $\Omega=0$
are displayed, it is seen that
most $\vec l$-vectors point to the negative $z$ direction
except that near the upper and lower surface region 
$\vec l$-vectors acquire $x$-$y$ component.
The boundary condition enforces  the $\vec l$-vector
parallel to the surface. It is especially strong when the
curvature of the surface is large.
Thus, in this case (see Fig. 5(a)) the left and right ends force the
$\vec l$-vectors point to $z$ direction, determining the overall
$\vec l$-vector configuration to the negative $z$ direction,
even in deep inside vectors.
The amplitude of the $\vec l$-vector decreases towards the outside.
At the center the $\vec l$-vector is completely locked to the negative $z$ direction
and towards the outside it becomes twisting, which ultimately
generates the bending mass current in $x$-$y$ plane as shown in Fig. 5(b).
However, this in-plane circular current diminishes further outside
because the absolute magnitude of the condensate decreases.
Therefore, the current maximum occurs in the intermediate circular region.

Under rotation, in stead of the MH vortex in the cigar case,
a pair of the Anderson-Toulouse-Chechetkin(ATC) vortex\cite{AT}  appears one by one
as increasing rotation.
This is understood because the overall $\vec l$-vectors tend to align 
to the negative $z$ direction from the outset, thus the
$\vec l$-vector of the core with the positive  $z$ direction 
changes over completely the surrounding $\vec l$-vector
with the negative $z$ direction. This object is nothing but ATC
vortex.

It might be interesting to point out that
the direction of the spontaneously generated mass current 
at rest is always perpendicular to the
direction of the majority $\vec l$-vectors, that is,
in the cigar (pancake) case the $\vec l$-vectors point to
$x$-$y$ ($z$), so the mass current flows to the $z$ ($x$-$y$) direction.
This implies that in our system the shape is decisive in understanding 
and controlling the physics of the textures.

In conclusion, we have found stable $\vec l$-vector textures
at rest and under rotation for $p$-wave chiral superfluid to be realized 
by using Feshbach resonances of neutral atom gases.
These textures are quite generic independent of the anisotropy
$\alpha$, provided that temperature is low enough, or 
the system is in the A region in Fig.1.
The spontaneous mass current and the $\vec l$-vector textures
can be used as diagnoses to detect and characterize $p$-wave superfluidity.

We thank T. Ohmi, M. Ichioka, T. Mizushima, and T. Kawakami for useful discussions.

\end{document}